\documentclass[prl,twocolumn,showpacs,superscriptaddress,floatfix]{revtex4}
\usepackage{graphicx}

\newcommand{\be}{\begin{equation}}
\newcommand{\ee}{\end{equation}}
\newcommand{\bea}{\begin{eqnarray}}
\newcommand{\eea}{\end{eqnarray}}

\newcommand{\f}{\frac}
\newcommand{\SQW}{\mbox{$S(\vec{q},\omega)$}}

\begin{document}

\title{Numerical evidence for unstable magnons at high fields in the Heisenberg antiferromagnet on the square lattice}
\author{Olav  F. Sylju{\aa}sen}
\affiliation{Department of Physics, University of Oslo, P.~O.~Box 1048 Blindern, N-0316 Oslo, Norway}

\date{\today}


\begin{abstract}
We find evidence for decaying magnons at strong magnetic field in the square lattice spin-$1/2$ Heisenberg antiferromagnet. The results are obtained using Quantum Monte Carlo simulations combined with a Bayesian inference technique to obtain dynamics and are consistent with predictions from spin wave theory.
\end{abstract}

\maketitle

The square lattice spin-$1/2$ Heisenberg antiferromagnet (2DHAF) is the archetype quantum antiferromagnet and describes the magnetism of the mother compounds of the high-Tc cuprate materials. While the 2DHAF itself is rather well understood, less is known when additional interactions are added. In this Rapid Communication we report on the nature of the excitations of the 2DHAF in an external magnetic field. Such a perturbation is highly relevant from an experimental point of view and has very interesting consequences.

Adding a magnetic field introduces interactions among the magnon excitations of the pure 2DHAF. According to spin wave theory these interactions will cause the magnons in certain regions of the Brillouin zone to be unstable at very high fields\cite{Zhitomirsky}.  One should however note that spin wave theory, which is an expansion in the parameter $1/S$, is not at all guaranteed to work for $S=1/2$. Furthermore the spin wave prediction involves certain assumptions as it is a self-consistent calculation which neglects vertex corrections and renormalizes the spin wave spectrum considerably. In addition there are also other theoretical predictions of what happens to the 2DHAF in a magnetic field: A scenario where the spin-1 excitation is viewed as a composite excitation of two dimensional spinons predicts additional bands of low-energy excitations at moderate values of the field\cite{Hsu,Palee}. Thus to settle the issue an independent calculation of the excitation spectrum of the 2DHAF in a magnetic field is needed. It is rather surprising that no such calculation has been performed until now, given the prominence and simplicity of the model. In this Rapid Communication we use Quantum Monte Carlo (QMC) simulations combined with a Bayesian inference technique to give a detailed and unbiased description of the excitation spectrum of the 2DHAF in a magnetic field. We find regions with broad spectral features at strong magnetic fields indicating decaying magnons consistent with spin wave theory, and we also show details of the spectrum in the decaying regions.

In zero magnetic field the ground state of the 2DHAF is N{\'e}el ordered with a renormalized moment. Spin wave theory explains very well the renormalized moment as well as the dispersion relation of the spin-1 excitations, except for an anomaly occurring at ($\pi$,$0$) (we use units where the lattice spacing $a=1$) where the dispersion softens and the spectrum broadens anomalously\cite{OlavRonnow,Sandviksingh,Ronnowexp}. This anomaly at $(\pi,0)$ is on the contrary natural in a picture of the spin-1 excitation as a composite excitation of two spin 1/2 spinons that are excitations about a mean-field pi-flux state\cite{Hsu}. While this bound state has almost the same dispersion as predicted by spin wave theory it differs around $(\pi,0)$ because the two spinons making up the bound state each has low energy close to the nodal pockets at $(\pm\pi/2,\pi/2)$. Furthermore this spinon picture can also explain the anomalous broadening of the continuum part of the spectrum at $(\pi,0)$\cite{Anderson}. Thus the spinon picture works well for explaining the $(\pi,0)$ anomaly qualitatively, however being essentially a projected mean field calculation it is not of the level of precision needed in order to compare quantitative details with exact results and conventional spin wave theory.
 
In a magnetic field the distinction between the spinon and magnon picture is more dramatic. By assuming that the spinon bands move rigidly when changing the magnetic field, as happens in the 1D Heisenberg antiferromagnet, the dynamic structure factor was calculated in Ref.~\cite{Palee}. The main result of this is the prediction of a new band of excitations forming an incommensurate ring at low energy close to $(\pi,\pi)$ at intermediate fields. In contrast, spin wave theory predicts a smooth change of the magnon dispersion with field up to a high value at which the magnons in some regions of the Brillouin zone become unstable and decay\cite{Zhitomirsky}.

The Hamiltonian of the 2DHAF in a magnetic field is
\be \label{Hamiltonian}
  H = J \sum_{\langle i,j \rangle} \vec{S}_i \cdot \vec{S}_j - B \sum_i S^z_i
\ee
where $\vec{S}_i$ are spin-1/2 operators, $\langle i,j \rangle$ indicates nearest-neighbor lattice sites on a square lattice, $J$ is the exchange constant and $B$ is the strength of the magnetic field, here pointing along the spin-z axis. 

This Hamiltonian can be simulated efficiently using QMC methods. Specifically we use the stochastic series expansion technique with directed-loop updates\cite{SS}. In order to measure the excitations of the system we will focus on calculating the dynamic structure factor which is the primary quantity measured in neutron scattering
\be
 S^{\alpha}(\vec{q},\omega) = \int_{-\infty}^{\infty} dt e^{i\omega t} \langle S^\alpha_{\vec{q}}(t) S^\alpha_{-\vec{q}}(0) \rangle
\ee
where $S^\alpha_{\vec{q}} = (1/N)\sum_{\vec{r}} S^\alpha_{\vec{r}}$ and $N=L^2$ is the total number of lattice sites. We use $L=32$ when not stated otherwise. The spin component direction is denoted by $\alpha$ and can be either $x,y$ or $z$. 

The dynamic structure factor is not directly accessible in QMC because the method is formulated in imaginary time. In contrast, what is accessible in the QMC simulations is the {\em imaginary-time} correlation function $D^{\alpha}(\vec{q},\tau)=\langle S^\alpha(\vec{q},\tau) S^\alpha(-\vec{q},0) \rangle$ which is related to the dynamic structure factor as
\be \label{imrealrelation}
    D^{\alpha}(\vec{q},\tau) = \int_0^{\infty} \f{d\omega}{2\pi} \left( e^{-\omega \tau} + e^{-(\beta-\tau)\omega} \right) S^{\alpha}(\vec{q},\omega).
\ee
A direct inversion of Eq.~(\ref{imrealrelation}) is difficult as the kernel has many near zero eigenvalues which causes the errors of $D$ to be amplified enormously. Instead we use a statistical inversion method known as the Average Spectrum Method(ASM)\cite{ASM} where one takes as the resulting structure factor the weighted average of all possible structure factors $S_{(i)}^\alpha(\vec{q},\omega)$ where the weight factor depends on how well the associated imaginary-time correlation function $D_{(i)}^\alpha$  matches the imaginary-time data $D^\alpha$. In practice this weighted average is computed using another Monte Carlo simulation, and details are available in Ref.~\cite{OlavASM}.

\begin{figure}
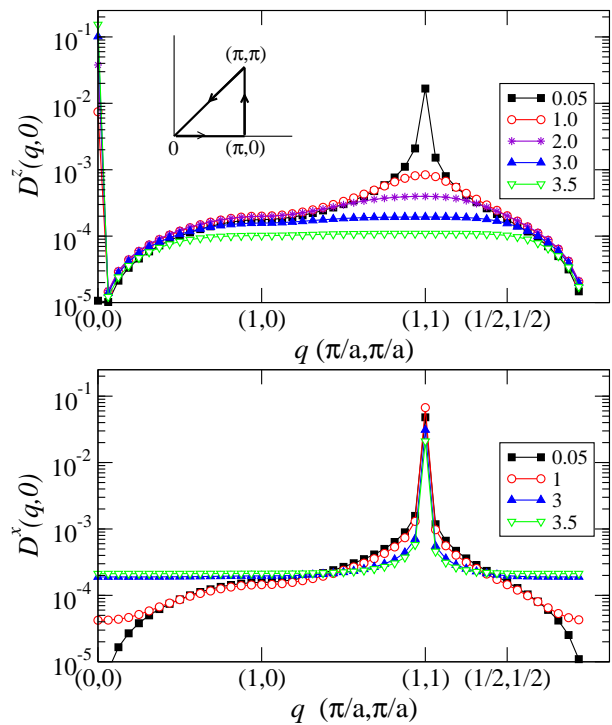

\includegraphics[clip,width=8cm]{fig1a}
\includegraphics[clip,width=8cm]{fig1b}
\caption{Instantaneous correlation function $D^{\alpha}(\vec{q},0)$ vs. $\vec{q}$ along the path in the Brillouin zone indicated in the upper left corner of the top panel. The different curves are for different values of the magnetic field $B$ specified in the legends. The upper panel is for longitudinal spin polarizations ($\alpha=z$) while the lower is for transverse polarizations ($\alpha=x$). The error bars are smaller than the symbol size. \label{static}}
\end{figure}

Before we present the results for the dynamic structure factor, we show in Fig.~\ref{static} the instantaneous spin-correlation function $D^{\alpha}(\vec{q},0)$ for longitudinal (along the magnetic field) and transverse spin components, respectively. Using Eq.~(\ref{imrealrelation}) these results translate into a measure of the total strength of excitations at a given $\vec{q}$-point.  
For the longitudinal spin component the effect of the magnetic field is to weaken the antiferromagnetic signal at $(\pi,\pi)$ and enhance the ferromagnetic signal at $(0,0)$ reflecting uniform canting of spins along the magnetic field. In contrast the antiferromagnetic signal at $(\pi,\pi)$ remains even at high magnetic fields for the transverse spin components.
For high fields note that $D^{\alpha}(\vec{q},0)$ becomes nearly independent of $\vec{q}$ except close to $\vec{q}=0$ and $\vec{q}=(\pi,\pi)$ for the longitudinal and transverse polarization respectively.   

\begin{figure}
\includegraphics[clip,width=8cm]{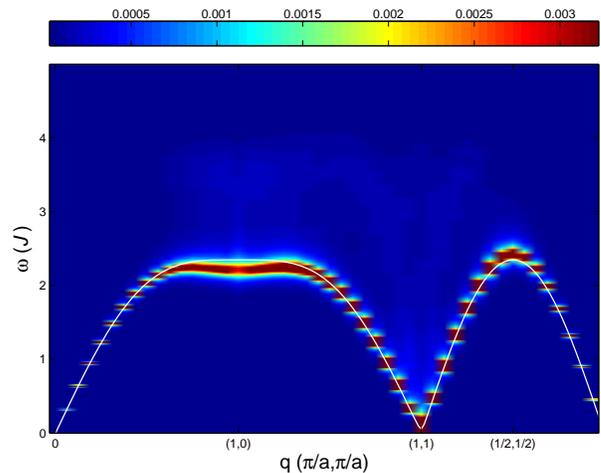}
\caption{Color-scale plot of $S^{z}(Q,\omega)$ where $q$ takes values along a path in the Brillouin zone specified in the inset. The thin white curve indicates the renormalized linear spin wave dispersion, $Z_c \omega_s(\vec{q})$. \label{smallfield}}
\end{figure}

In a weak magnetic field we expect that the dynamic structure factor is essentially the same as in zero field. Fig.~\ref{smallfield} shows a color-scale plot of $S^{z}(\vec{q},\omega)$ along the same scan in $\vec{q}$-space as in Fig.~\ref{static} for $B/J=0.05$ at $T=J/40$. The color-plot is made by discretizing the frequency interval $\omega/J=[0,8]$ into 1000 bins and assigning a color to the value of $S(\vec{q},\omega)$ within each interval. If the value of $S(\vec{q},\omega)$ corresponds to a value higher than the maximum color value, the color is set to the maximum color value. 

In accordance with the zero-field result\cite{Sandviksingh} we find that the magnon energy is lower at $(\pi,0)$ than at $(\pi/2,\pi/2)$. More precisely, the location of the peaks in $\SQW$ at the two values of momenta are $\omega_{\rm peak,(\pi,0)}=2.21 \pm 0.02$ and $\omega_{\rm peak,(\pi/2,\pi/2)}=2.42 \pm 0.02$, corresponding to a $9\%$ difference. While the dispersing magnon peak is the dominant feature of Fig.~\ref{smallfield} it is also possible to get a glimpse of high energy features. Noteworthy is the broadness of the magnon peak at $(\pi,0)$ and the V-shaped feature above $(\pi,\pi)$.

\begin{figure}
\includegraphics[clip,width=8cm]{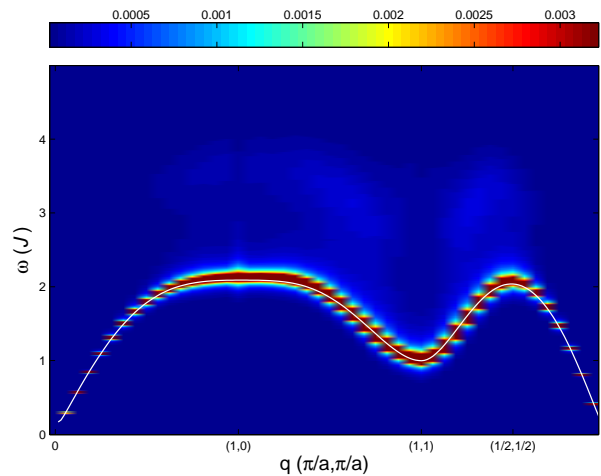}
\caption{Color-scale plot of $S^{z}(\vec{q},\omega)$ along the $\vec{q}$-path indicated in \ref{static}. The thin white curve is the perturbative result for the spin wave dispersion using the self-energy. \label{intermediatefield}}

\end{figure}

In a weak magnetic field the spins will orient themselves in a staggered configuration almost perpendicular to the field. As the field increases the spins will cant more and more along the field, and finally at $B=4J$ will align completely with the external magnetic field. From a symmetry perspective a magnetic field will break the SU(2) spin symmetry of the 2DHAF. This breaking of a global symmetry is reflected in the spectrum by the fact that one of the two goldstone modes gets a mass. In the longitudinal channel this manifests itself at $(\pi,\pi)$ where the zero-energy mode acquires a gap equal to the value of the magnetic field. In the transverse channel it is the $\vec{q}=0$ mode that becomes gapped. Fig.~\ref{intermediatefield} shows the longitudinal dynamic structure factor for $B/J=1$. 

The dispersion seen in Fig.~\ref{intermediatefield} agrees quantitatively with spin wave theory. By parameterizing the Hamiltonian Eq.~(\ref{Hamiltonian}) in a canted coordinate system, characterized by the canting angle $\theta$, and expressing the spin operators by boson operators according to the Holstein-Primakoff transformation one gets terms with all orders of bosonic creation and annihilation operators, including linear and cubic terms. Minimizing the energy of the zeroth order term gives a condition relating the canting angle to the magnetic field: $\sin{\theta}= B/4J$. With this value of $\theta$ the linear term vanishes identically, and the quadratic term can be diagonalized, yielding 
 \be \label{baredispersion}
  \omega^0_{\vec k}= 2J \sqrt{(1-\gamma_{\vec k})(1+\cos 2\theta \gamma_{\vec k})}
\ee
where $\gamma_{\vec{k}}= (\cos{k_x}+\cos{k_y})/2$.
There are corrections to this dispersion coming from higher order terms. The 3rd and 4th order terms in the expansion of the Hamiltonian are
\be \label{Hamilton3}
  H^{(3)}= - J \sum_{\langle i,j \rangle} \sqrt{\f{S}{2}} \sin(2\theta) \left( (a_i+a^\dagger_i)n_j + (i \leftrightarrow j) \right)
\ee
\bea 
  H^{(4)} &=& \f{J}{4} \sum_{\langle i,j \rangle}  \left\{ -2\cos 2\theta \; n_i n_j +\sin^2 \theta \; a_i^\dagger \left( n_i+n_j \right) a_j \right. \nonumber \\
    &  & \left. -\cos^2 \theta \; a_i^\dagger \left( a_i a_i + a_j^\dagger a_j^\dagger \right)) a_j + (i \leftrightarrow j) \right\} \label{Hamilton4}
\eea
Following Ref.~\cite{Zhitomirsky} we find corrections to the dispersion Eq.~(\ref{baredispersion}) by calculating the self-energy $\Sigma$ by contracting two cubic terms and retaining only the frequency independent parts of the quartic term. The perturbative correction to the magnon dispersion is then
\be \label{renormalizedspinwave}
   \omega_{\vec k} = \omega^0_{\vec k} + \Sigma(\omega^0_{\vec k},\vec{k}).
\ee
In zero field this reduces to a purely multiplicative renormalization $\omega_{\vec k} = Z_c \omega^0_{\vec k}$, where $Z_c=1.1765$ \cite{Canali} and is seen as a white curve in Fig.~\ref{smallfield} while the white curve in  
Fig.~\ref{intermediatefield} indicates $\omega_{\vec{k}}$ according to Eq.~(\ref{renormalizedspinwave}) for $B=J$.

Fig.~\ref{intermediatefield} reveals also that the broad peak seen at $(\pi,0)$ is also present at $B/J=1$. There are no appreciable continuum above $(\pi,\pi)$, but high energy features are seen going away from $(\pi,\pi)$.  

\begin{figure}
\includegraphics[clip,width=8cm]{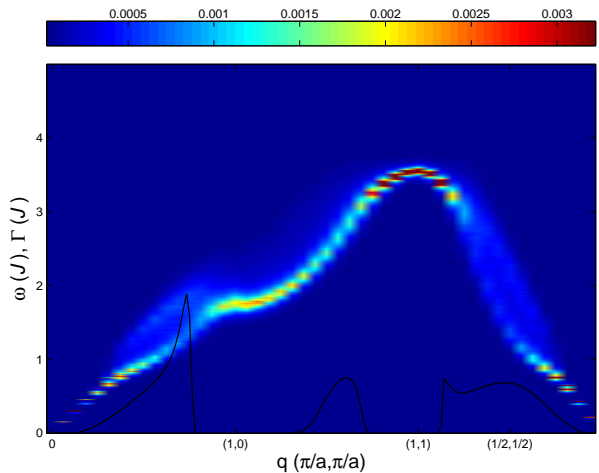}
\caption{Color-scale plot of $S^{z}(\vec{q},\omega)$ along the $\vec{q}$-path indicated in Fig.~\ref{static} for $B/J=3.5$. The thin black curve is the magnon linewidth $\Gamma$ calculated using the golden rule Eq.~(\ref{GoldenRule}). \label{largefield}}
\end{figure}

When the magnetic field is increased further our results show that the dynamic structure factor is still dominated by a large dispersing magnon peak with a gap at $(\pi,\pi)$ that continues to increase. This holds until the field reaches $B \approx 3J$, where the magnetization per site is $M=0.318$, i.e. about 63\% of the saturation value, at which the peaks at q-space points in the region around $(\pi/2,\pi/2)$ broadens and become less well-defined. Fig.~\ref{largefield}, which is the main result of this Rapid Communication, shows the dynamic structure factor at a field $B/J=3.5$ ($M=0.392$) where these effects reach a maximum. 

\begin{figure}
\includegraphics[clip,width=8cm]{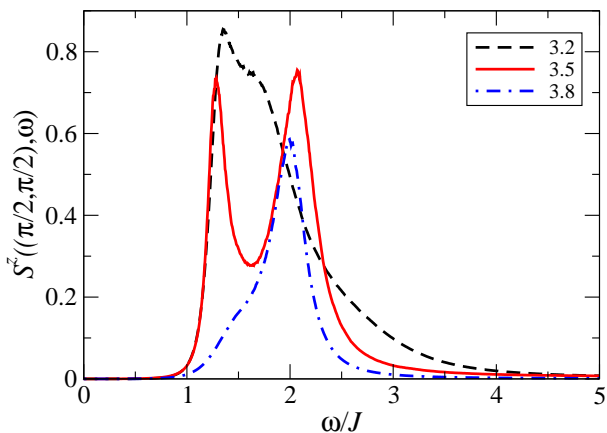}
\caption{(Color online) Lineshapes of the dynamic structure factor at $(\pi/2,\pi/2)$ for three different values of magnetic field as indicated in the legend. Here the system size is $L=16$ and the inverse temperature is $\beta J=200$.\label{lineshapes_highfields}}
\end{figure}
For magnetic fields above $H \approx 3J$ the self-energy $\Sigma$ acquires an imaginary part due to decay of magnons thru the three-boson vertex in Eq.~(\ref{Hamilton3}). This implies that the perturbation expression Eq.~(\ref{renormalizedspinwave}) will not work well for finding the dispersion and a more sophisticated analysis is needed\cite{Zhitomirsky}. An estimate for the imaginary-part of the self-energy, or equivalently the magnon linewidth, can however simply be obtained using Fermi's Golden rule
\be \label{GoldenRule}
  \Gamma = \f{2\pi}{\hbar} \sum_{\vec q} |\langle {\vec k}-{\vec q},{\vec q} | H^{(3)} | {\vec k} \rangle|^2 \delta(\omega^0_{\vec{k}-\vec{q}}+\omega^0_{\vec{q}} - \omega^0_{\vec{k}}). 
\ee 
where we use the bare spin-wave dispersion. The result of this is plotted in Fig.~\ref{largefield} and agrees qualitatively with the location of the decaying regions. This result depends mostly on the available phase space, i.e. the spin-wave dispersion, as the interaction vertex is a slowly varying function of $\vec{q}$.

Note that the decay of a spin-1 excitation into two spin-1 excitations is not forbidden by the symmetry of the Hamiltonian. Not even at $B=0$ where the Hamiltonian has a full SU(2) symmetry. However, at $B=0$ there are no terms in the Hamiltonian with an odd number of boson terms, thus a single magnon can only decay into an odd number of  magnons. In contrast cubic terms are present in a magnetic field.

Fig.~\ref{largefield} reveals a double-peaked structure in the decaying regions. Looking more closely at this we show in Fig.~\ref{lineshapes_highfields} lineshapes at $(\pi/2,\pi/2)$ for different values of the magnetic field. For the $B/J=3.5$ curve one can clearly identify two peaks, one at $\omega \approx 1.28J $ and one at $\omega \approx 2.06J$. However these peaks are overlapping. This could be either because the ASM cannot resolve the peaks completely or that the spectrum is really a continuum where the two peaks indicate enhanced density of states near the edges. We favor the latter interpretation as the temperature is much less than the spacing between the peaks and we have collected very accurate QMC data (std.dev. $\sim 1 \times 10^{-6}$). This should in our experience be enough to resolve the peaks if they were distinct\cite{OlavASM}. In view of the continuum interpretation the effect of increasing the magnetic field in this region is then to transfer weights from the lower to the upper edge of the continuum.  

In no circumstances do we see ring-like regions of low-energy excitations as predicted by the spinon bound state scenario. Thus our numerical results do not favor this interpretation.

The magnon decay reported here is seen at very high magnetic fields. Experimentally this decay is therefore most relevant in materials with a low value of the exchange constant\cite{Landee}, and perhaps also for systems with cold fermions in an optical lattice\cite{ColdGases}.

The simulations were carried out using the Titan cluster at the University of Oslo and the Nordugrid data production facility through the Nordugrid ARC middleware.

\end{document}